\documentclass[preprint,5p,number,a4paper,sort&compress]{elsarticle}
\usepackage{graphicx}
\usepackage{epstopdf}
\usepackage{dcolumn}
\usepackage{amsmath}
 \usepackage{amssymb}
\usepackage{courier, pifont}
\usepackage[scaled=0.92]{helvet}
\usepackage[T1]{fontenc}
\usepackage{textcomp}
\usepackage{color}
\usepackage[displaymath, mathlines]{lineno}
\usepackage{cases}
\modulolinenumbers[1]

\newcommand{\red}{\textcolor{black}}
\newcommand{\blue}{\textcolor{black}}
\journal{Nuclear Instruments and Methods in Physics Research Section A}

\let\oldequation\equation
\let\oldendequation\endequation
\renewenvironment{equation}
  {\linenomathNonumbers\oldequation}
  {\oldendequation\endlinenomath}
\begin{document}
\begin{frontmatter}
\title{Particle identification and revolution time corrections for the isochronous mass spectrometry in storage rings}
\cortext[mycorrespondingauthor]{Corresponding authors}
\author[A,B]{Y.M. Xing}
\author[A,C,D]{Y.H. Zhang\corref{mycorrespondingauthor}}
\ead{yhzhang@impcas.ac.cn}
\author[A,C,D]{M. Wang\corref{mycorrespondingauthor}}
\ead{wangm@impcas.ac.cn}
\author[A,B]{Yu.A. Litvinov\corref{mycorrespondingauthor}}
\ead{y.litvinov@gsi.de}
\author[A,B]{R.J. Chen}
\author[A]{X.C. Chen}
\author[A]{C.Y. Fu}
\author[A,C]{H.F. Li}
\author[A]{P. Shuai}
\author[A,C]{M. Si}
\author[A]{M.Z. Sun}
\author[A,C,D]{X.L. Tu}
\author[A,C]{Q. Wang}
\author[A,C,D]{H.S. Xu}
\author[A,E]{X. Xu}
\author[A]{X.L. Yan}
\author[A,C]{J.C. Yang}
\author[A,C]{Y. J. Yuan}
\author[A,F]{Q. Zeng}
\author[A,C]{P. Zhang}
\author[A,C]{M. Zhang}
\author[A,C]{X. Zhou}
\author[A,C,D]{X.H. Zhou}
\address[A] {Key Laboratory of High Precision Nuclear Spectroscopy and Center for Nuclear Matter Science, Institute of Modern Physics, Chinese Academy of Sciences, Lanzhou 730000, China}
\address[B] {GSI Helmholtzzentrum f{\"u}r Schwerionenforschung, Planckstra{\ss }e 1, 64291 Darmstadt, Germany}
\address[C] {School of Nuclear Science and Technology, University of Chinese Academy of Sciences, Beijing, 100049, China}
\address[D] {Joint Research Center for Modern Physics and Clean Energy, South China Normal University, Institute of Modern Physics CAS, China}
\address[E] {School of Science, Xi'an Jiaotong University, Xi'an, 710049, China}
\address[F] {School of Nuclear Science and Engineering, East China University of Technology, NanChang, 330013, China}

\begin{abstract}
In the isochronous mass spectrometry (IMS) performed at storage rings, masses of short-lived nuclides are determined through precision measurements of their mean revolution times. However, the distribution of revolution times could be seriously deteriorated by instabilities of \red{the} ring's magnetic fields. This becomes a significant obstacle for the particle identifications and mass determinations. A data analysis method is described in this paper which is able to largely remove the uncertainties caused by the \red{magnetic} field instabilities in the particle identifications and the mean revolution times. We show that this method is very effective for the IMS experiments even when the \red{magnetic} fields of a storage ring vary slowly up to a level of $\Delta B/B\sim 10^{-4}$.

\end{abstract}

\begin{keyword}
\texttt ~Isochronous mass spectrometry\sep Storage ring\sep Particle identification\sep Revolution time correction
\end{keyword}

\end{frontmatter}

\section{Introduction}
{\red{Nuclear binding energy is a fundamental characteristic of an atomic nucleus which reflects the net effect of complex interactions between its composing nucleons.
Nuclear binding energies are straightforwardly derived from nuclear masses.
The latter are determined experimentally.
In nuclear structure, precision mass values of radioactive nuclei are used to investigate shell structure, limits of nuclear existence, nucleon-nucleon correlations and many other effects.
Nuclear masses are an indispensable input to astrophysical models of nucleosynthesis processes in the universe.
Thanks to the broad range of applications of nuclear masses, nuclear mass spectrometry is an intensively developing field of modern experimental physics~\cite{Blaum2006,Sun2018}.
Mass measurements are conducted (planned) at almost all presently operated (future) radioactive ion beam facilities.
The challenge today is to achieve measurements of very exotic nuclei which are produced with small cross-sections and have short lifetimes.}}

Isochronous Mass Spectrometry (IMS)~\cite{HausmannNIMA2000} employed at heavy-ion storage rings is an efficient tool for precision mass measurements of short-lived nuclides.
%
The IMS technique relies on the accurate determination of the revolution times of relativistic secondary nuclides stored in an isochronously tuned storage ring.
The latter setting means that the revolution times of the particles should (in first order) not depend on their momenta.
Details about the IMS technology can be found in Refs.~\cite{Sun2018,Zhang2016}. Here we only give a brief introduction.

Several IMS experiments have been performed successfully in the storage rings CSRe (Cooler Storage Ring for experiment) at Institute of Modern Physics (IMP)~\cite{TuPL2011,ZhangPRL2012,YanAPL2013,ShuaiPLB2014,XuPRL2016,ZhangPLB2017,Zhang2018,Xing2018,Fu2018} and ESR (Experimental Storage Ring) at \red{GSI Helmholtz Center for Heavy Ion Research (GSI)}~\cite{Stadlmann2004,Sun2008,Sun2010,Knobel2016A,Knobel2016B}.
In those experiments, projectile fragments within a certain momentum acceptance were injected and stored in a storage ring.
Most of the fragments were fully-striped nuclei.
Their revolution times, $T$, were measured using dedicated time-of-flight detectors~\cite{Tr92,Meib2010,Zhangw2014}.

In principle, the mass-to-charge ratio, $m/q$, of an ion circulating in the ring is determined by the expression:
  \begin{equation}\label{MvQ}
  m/q=B\rho/(\gamma v)=B\rho\sqrt{\frac{T^2}{C^2}-\frac{1}{v_c^2}},
  \end{equation}
where $\gamma$ and $v$ are the relativistic Lorentz factor and velocity of the ion. $B\rho$ and $C$ are the ion's magnetic rigidity and the orbital length respectively, and $v_c$ is the speed of light in vacuum. Due to technical limitations, only $T$ of the stored ions can be precisely measured in IMS experiments. The unknown masses of the nuclides are determined via the $m/q$ versus $T$ calibration using the nuclides with precisely known masses. This requires that $T$ for each ion species is determined with high accuracy.

Compared with the performance of ToF detectors used in the IMS experiments, two major factors contribute to the uncertainty of $T$ determinations: (i) the non-zero momentum acceptance of the storage ring and (ii) the instabilities of the ring's magnetic fields during measurements. The influence of the betatron motion of nuclides on $T$ is usually canceled out by averaging of multi-turn detections.

Firstly, the momentum acceptance of a storage ring for IMS experiments is typically in the order of $10^{-3}$, meaning that the variation of $B\rho$ or $C$ is also of the same level.
This results in obvious spreads of $T$.
Fortunately, in an isochronously tuned storage ring, a larger $B\rho$ corresponds to a larger $C$, and vice versa.
This tends to make $T$ less dispersive according to Eq.~(\ref{MvQ}).
The final spreads of $T$ due to the finite acceptance of a storage ring can be expressed as~\cite{HausmannNIMA2000}
\begin{equation}\label{DeltaT_Bp}
\delta T=\left(\frac{1}{\gamma_t^2}-\frac{1}{\gamma^2}\right)\frac{\delta (B\rho)}{B\rho}T,
\end{equation}
or
\begin{equation}\label{DeltaT_C}
\delta T=\left(1-\frac{\gamma_t^2}{\gamma^2}\right)\frac{\delta C}{C}T,
\end{equation}
where $\delta (B\rho)/(B\rho)$ and $\delta C/C$ represent the corresponding $B\rho$ and $C$ variation of ions in a storage ring, respectively. $\gamma_t$ is the so-called transition energy of a storage ring, which is fixed during the whole experiment.
According to Eq.~(\ref{DeltaT_Bp}) and Eq.~(\ref{DeltaT_C}), the spread of $T$ due to the non-zero momentum acceptance is dramatically reduced for the ions with $\gamma$ equal or close to $\gamma_t$, i.~e., the isochronous condition is fulfilled for ions with $\gamma\sim\gamma_t$. Usually, $\delta T$ varies in the range of $10^{-5}$ $T$ to $10^{-6}$ $T$ for different ions depending on the difference of their $\gamma$ from $\gamma_t$.

Secondly, if the ring dipole magnetic field $B$ changes by $\Delta B$, ions in front of the ring with different mean magnetic rigidity would be selected, injected and stored. The ring working with a different field setting shifted by $\Delta B$ would lead to a shift of $T$, $\Delta T$, estimated by
\begin{equation}\label{DeltaT_Xi}
  \Delta T=-\frac{1}{\gamma^2}\frac{\Delta B}{B}T.
 \end{equation}
$\Delta T/T$ has the same order of magnitude as $\Delta B/B$.
It has been observed that the magnetic fields can vary slowly up to a level of $\Delta B/B \sim 10^{-4}$. $\Delta T$ is usually much larger than $\delta T$ although $\Delta B/B$ is smaller than the $B\rho$ acceptance of the ring.

Since instabilities of magnetic fields considerably contribute to the deterioration of the $T$ resolution, the corresponding $T$ shifts $\Delta T$ should be corrected~\cite{Zhang2018}. 
Although the efforts to reduce the effects of field instabilities have never been ceased since the very first IMS experiments, there is still room for further improvements~\cite{shuai-MFC}.

Given the fact that each measurement for a single injection lasts for less than one millisecond, the magnetic fields can safely be assumed to be stable within this short period of time. If ions are identified in this injection, the information extracted from the precisely known $m/q$ values of the ions can be used to correct their $T$. In the following, we introduce a data analysis technique for particle identification (PID) in a single injection, then we present a new method for $\Delta T$ corrections on the injection-by-injection basis. For a better understanding, we use the data-set from a recent experiment conducted at the HIRFL-CSR accelerator complex in IMP, Lanzhou.
In this experiment, $^{112}$Sn was used as the primary beam.
The storage ring CSRe was set to $B\rho$=5.3347 Tm and $\gamma_t=1.302$.
The instabilities of magnetic fields during the experiment were at the level of $10^{-4}$.
The estimated momentum acceptance was $\Delta{p}/p\approx0.9\times10^{-3}$ (FWHM).
More details of the experiment can be found in Ref.~\cite{Xing2018}.
\blue{The preliminary results reported in Ref.~\cite{JLiu} were obtained by using the method described in this work.}

\section{Particle identification (PID)}
\subsection{Conventional PID method and its limitations}\label{CPID}
\begin{figure}[hbt]
	\centering
	\includegraphics[angle=0,width=8.5 cm]{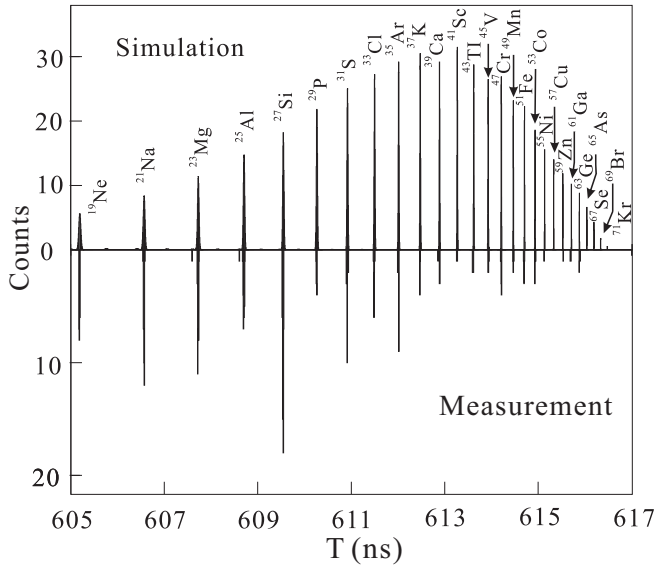}
	\caption{An example of the PID method based on the matching of the simulated and measured $T$ spectra (taken from Ref.~\cite{Tu2011NIMA}). A prerequisite in this PID method is that the $T$ peaks corresponding to different nuclides in the measured $T$ spectrum should be well resolved. The count differences between the calculated and measured peaks may be due to the odd-even staggering in production yields, which are not taken into account in simulations~\cite{Mei2014} as well as due to different storage times and detection efficiencies for heavy and light ions. We note, however, that the count differences do not affect the PID method.
	\label{Tu_PID}}
\end{figure}

The $T$ values of stored ions in an individual injection were extracted in the same way as in Ref.~\cite{Tu2011NIMA}. The flight times of ions in the ring were recorded turn by turn for a maximum of about 300 turns and $T$ were determined through a third order polynomial fit of the total flight times as a function of the number of turns. Conventionally, these $T$ are put into a histogram combined from all injections, forming an integrated experimental $T$ spectrum as shown in the lower panel of Fig.~\ref{Tu_PID}. The peaks in such a $T$ spectrum we name as "$T$ peaks" in the following discussions. For the $N$ well-resolved $T$ peaks in Fig.~\ref{Tu_PID}, their mean $T$s were deduced and labeled as $\overline{T}_{exp}^i$ with $i=1,2, ... N$.

The $T$ spectrum can be simulated~\cite{Matos2004} following
\begin{equation}\label{Tsim}
T=C{\sqrt{\left(\frac{m/q}{B\rho}\right)^2+\frac{1}{v_c^2}}}.
\end{equation}
Usually, one uses a set of nuclides with production cross sections and transmission predicted by LISE++ code~\cite{Lise1,Lise2} and their $m/q$ values provided by the Atomic Mass Evaluation, AME$^{\prime}$16~\cite{AME2016}. Then, by fixing $C$ as the central orbital length of the storage ring (128.80 m for CSRe~\cite{XiaNIMA2002}), a number of simulated spectra can be produced by adjusting the $B\rho$ values.
In fact, when changing the $B\rho$ value, the whole spectrum shifts on the $T$ axis, and the simulated $T$ peak intervals also change. Finally, one can find a simulated spectrum which gives a minimum value of $RMS$ (the root-mean-squares) defined as
\begin{equation}\label{RMS-1}
RMS=\sqrt{\sum_{i = 1}^N \frac{(\overline{T}_{exp}^i-T_{sim}^i)^2}{N}},
\end{equation}
where $T_{sim}^i$ is the simulated mean $T$ of a $T$ peak in the nearest proximity of $\overline{T}_{exp}^i$.
The best-matched simulated spectrum is illustrated in the upper panel of Fig.~\ref{Tu_PID} (such a process is called hereafter the peak {\it matching} procedure).
Since the ion species have already been assigned to each peak in the simulation, PID is automatically realized once the peak {\it matching} is achieved.

The PID method according to the peak {\it matching} procedure in the integrated spectrum is not sensitive to the $m/q$ values within reasonable uncertainties used in the simulations. This indicates that an ion with unknown mass can still be identified by using the estimated mass value from AME$^{\prime }$16~\cite{AME2016}. For instance, a relative variation of $T$ caused by the mass uncertainty, $\Delta m/m$, can be estimated according to
\begin{equation}\label{DTT}
\frac{\Delta T}{T}=\frac{1}{\gamma^2}\frac{\Delta m}{m}.
\end{equation}
For the isochronous mass measurements of $40\le A\le 100$ nuclei in CSRe, the time differences due to mass uncertainty of 1 MeV are less than 10 ps, which is much smaller than most of the intervals between $T$ peaks shown in Fig.~\ref{Tu_PID}.

\begin{figure}[hbt]
	\centering
	\includegraphics[angle=0,width=8.5 cm]{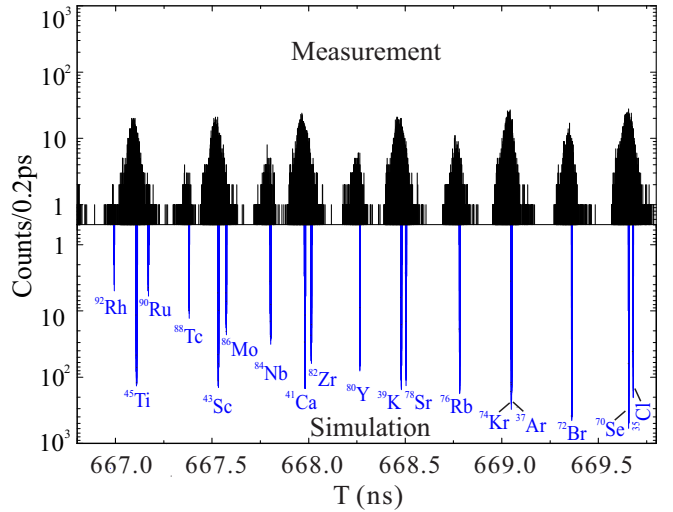}
	\caption{Part of the measured $T$ spectrum (up) in the experiment using a primary $^{112}$Sn beam and the corresponding simulated $T$ spectrum (bottom).
	It is difficult to unambiguously assign PID to all particles because of the overlap of the measured $T$ peaks due to the significant instabilities of the magnetic fields and small intervals between $T$ peaks.
		\label{Spectrum_Ori_Sim}}
\end{figure}

However, this conventional PID method works only when the $T$ peaks from different nuclides in the measured spectrum are well separated from each other. If the peaks are too broad due to the field instabilities, some peaks will overlap as shown in Fig.~\ref{Spectrum_Ori_Sim}. If the overlaps are frequently encountered in the integrated $T$ spectrum, it is quite difficult to use this method to realize PID.

Since reliable PID is an indispensable step of the overall analysis, we developed a new technique that can realize PID for each individual injection. This allows us to make the $\Delta T$ corrections on injection-by-injection basis.

\subsection{PID in a single injection}\label{PID2}
\begin{figure}[hbt]
	\centering
	\includegraphics[angle=0,width=8 cm]{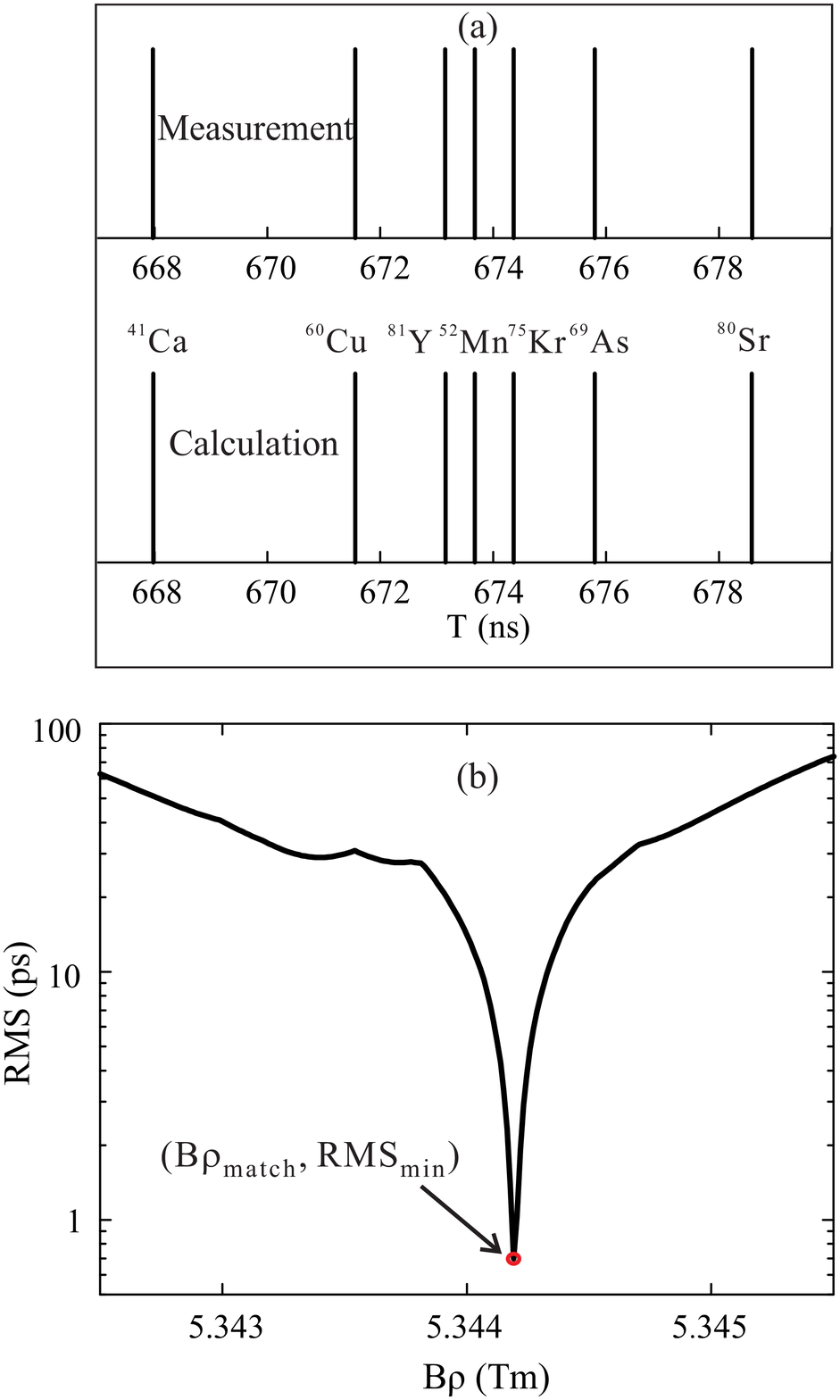}
	\caption{(a) The $matching$ procedure between the measured and calculated $T$ in a single injection at ($B\rho_{match},RMS_{min}$). (b) The variation of $RMS$ versus $B\rho$. The red circle with minimum $RMS$ indicates the converged $matching$ that yields PID in this injection.
		\label{PID-RMS-10039}}
\end{figure}

Assuming that $M$ ions are injected and their $T$ are measured to be $T_{exp}^i$ with $i=1,2, ... M$ ($M$ is limited to be less than about 40 in typical IMS experiments~\cite{ZhangNIMA2014}). \red{The} upper panel of Fig.~\ref{PID-RMS-10039}(a) shows an example of seven $T$ values measured in a single injection.

\begin{figure}[hbt]
	\centering
	\includegraphics[angle=0,width=8.5cm]{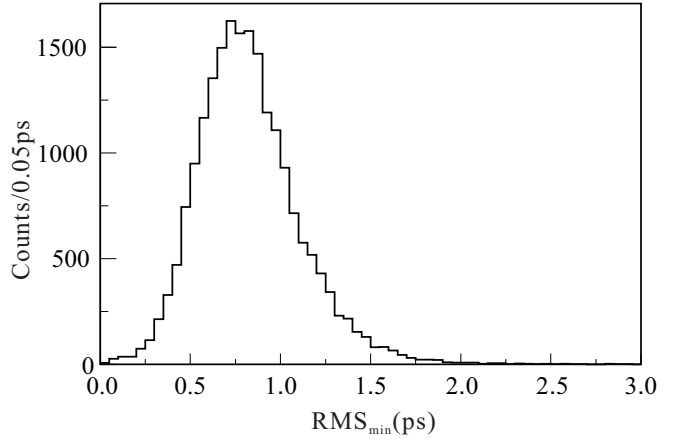}
	\caption{Distribution of $RMS_{min}$ from the $matching$ procedure applied to all injections.
		\label{RMS_min}}
\end{figure}

\begin{figure*}[hbt]
	\centering
	\includegraphics[angle=0,width=16 cm]{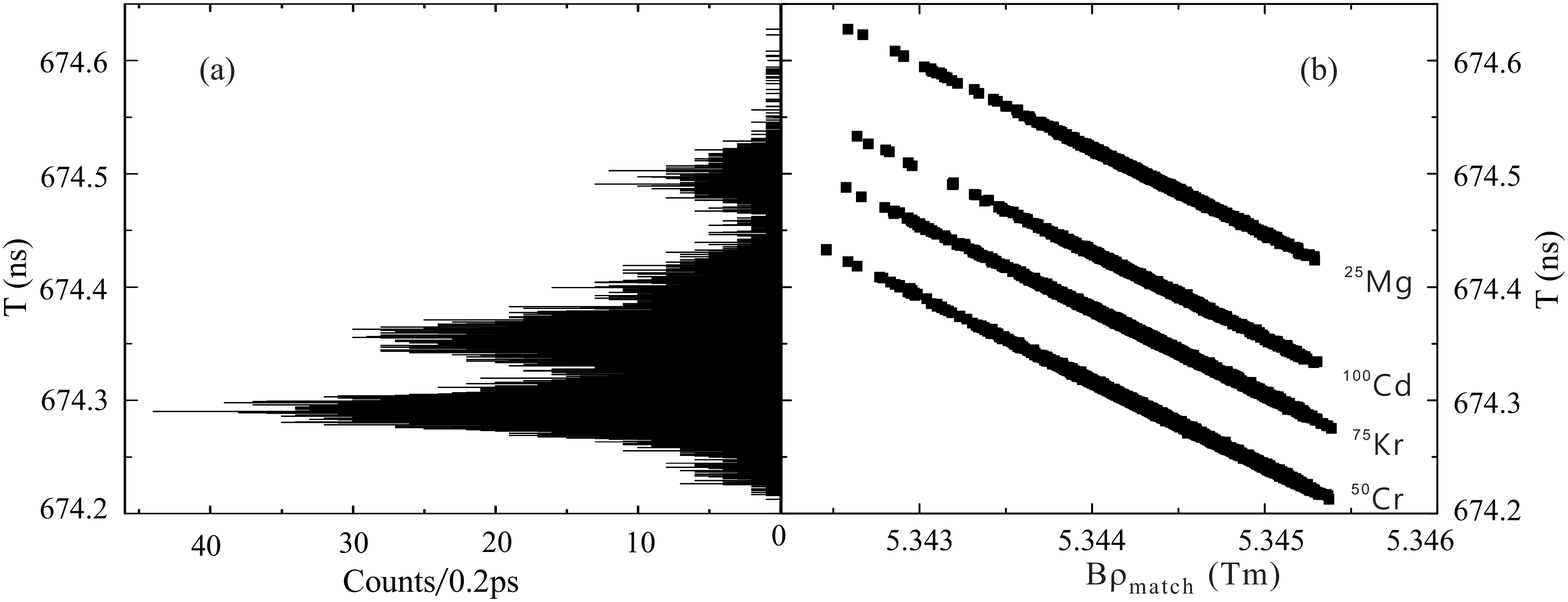}
	\caption{(a) A zoomed experimental $T$ spectrum at the range of 674.20 ns to 674.65 ns. The mixture of the peaks from four different ion species makes the PID very difficult. (b) The plot of measured $T$ versus $B\rho_{match}$ from which clear separations are observed.}
	\label{T-Brho-100Cd}
\end{figure*}

As mentioned in the previous section, a total of $N_t$ $T$, corresponding to $N_t$ ion species, can be calculated according to Eq.~(\ref{Tsim}) with a fixed $B\rho$ and central orbital length of $C=128.80$ m. They are labeled as $T_{cal}^i$ with $i=1,2, ... N_t$. Typically,
$N_t$ is much larger than $M$. One selects a calculated $T$, $T_{cal}^i$, in the nearest proximity of each $T_{exp}^i$, then the quantity defined as
  \begin{equation}\label{RMS-2}
  RMS=\sqrt{\sum_{i = 1}^M \frac{(T_{exp}^i-T_{cal}^i)^2}{M}}.
  \end{equation}
is calculated. 
Through scanning the $B\rho$ value in a reasonable range (in this case from 5.3425 Tm to 5.3455 Tm), the minimum $RMS$, $RMS_{min}$ can be found, \red{see, e. g., Fig.~\ref{PID-RMS-10039}(b)} \blue{and Fig.~3 in Ref.~\cite{JLiu}.}
\red{Once $RMS_{min}$ is uniquely determined, the experimental $T_{exp}^i$ matches the corresponding calculated $T_{cal}^i$, giving automatically PID.}
\red{If $M = 1$ or 2, there may be more than one local $RMS_{min}$ and an unambiguous automatic PID might be impossible.
Such cases need to be inspected separately and omitted if, e.g., unsolvable by combining (several) subsequent injections.}

Such analyses are applied to all injections, and the distribution of $RMS_{min}$ is shown as histogram in Fig.~\ref{RMS_min}. $RMS_{min}$ is a measure of the discrepancy between the experimental and calculated $T$. One sees that all $RMS_{min}$ values are smaller than 3 ps which is much smaller than the majority of intervals between different simulated $T$ peaks, as shown in Fig.~\ref{Spectrum_Ori_Sim}.
\begin{figure}[htb]
	\centering
	\includegraphics[angle=0,width=8.5 cm]{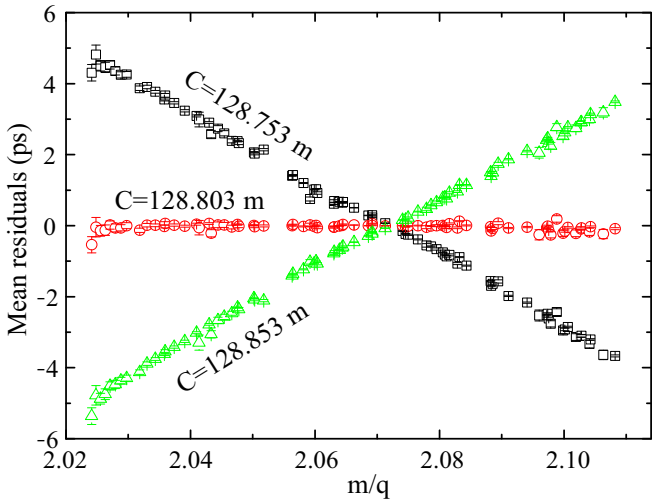}
	\caption{(Colour online) The mean residuals for different nuclides obtained in the matching procedure. Three cases with different $C$ values are shown for comparison. 
		\label{DeltaT-T-New}}
\end{figure}

The $B\rho_{match}$ value determined in the $matching$ procedure (see Fig.~\ref{PID-RMS-10039}(b)) is an approximation of the mean magnetic rigidity of the ions stored in this injection. It can also be used to demonstrate the reliability of PID.
Fig.~\ref{T-Brho-100Cd} shows a zoomed $T$ spectrum together with the scatter plot of the experimental $T$ versus $B\rho_{match}$ determined for each injection. One sees that the un-resolved $T$ in Fig.~\ref{T-Brho-100Cd}(a) can be clearly identified in Fig.~\ref{T-Brho-100Cd}(b) and thus unambiguously assigned to a specific ion species. For example, an ion with $T_{exp}=674.35$ ns can be identified as $^{50}$Cr or $^{75}$Kr in Fig.~\ref{T-Brho-100Cd}(a), but its definite assignment can be obtained in Fig.~\ref{T-Brho-100Cd}(b) with help of $B\rho_{match}$ value. Fig.~\ref{T-Brho-100Cd}(b) also shows that the relative variation of $B\rho_{match}$ is about $5.5\times 10^{-4}$, roughly reflecting the magnitude of the relative instability of magnetic fields. The $T$ values for one ion species vary in a range of 200 ps, which is mainly due to the field instabilities of the ring at the same level as $B\rho_{match}$.

Since the value of $C$ has been fixed to be 128.80 m in the {\it matching} procedure, the sensitivity of this choice has to be checked. Therefore, we vary $C$ within $C\pm 5$ cm in the {\it matching} procedure.
We calculated the residuals, $\sigma T_i=T_{exp}^i-T_{cal}^i$, for each ion in each injection, and then averaged them over all injections. Fig.~\ref{DeltaT-T-New} shows the mean residuals for all ion species. One sees that the mean residuals, $\overline{\sigma T_i}$, are scattered around zero when the orbit length $C=128.803$ m is used, while $\overline{\sigma T_i}$ deviates from zero if $C$ is different from 128.803 m by $\pm 5$ cm.
Note that the largest deviations are still less than 6 ps, so the {\it matching} procedure and consequently the PID are weakly sensitive to the initial choice of the orbital length $C$.

From the analyses described above, we have the following conclusions: (1) the central orbital length $C=128.80$ m~\cite{XiaNIMA2002} is a good approximation in calculating the $T_{cal}^i$ values according to Eq.~(\ref{Tsim}); (2) the $matching$ procedure as illustrated in Fig.~\ref{PID-RMS-10039} can be applied for the PID of the stored ions in a single injection; (3) once the minimum $RMS_{min}$ is found, we consider that the best $matching$ between the measured and calculated $T$ values has been achieved, giving automatically the PID for this injection.

The accuracy of this PID technique relies on the $T$ difference between ion species with similar $m/q$ ratios.
In IMS experiments in CSRe, definite PID can be achieved if the $T$ differences are larger than about 10 ps.
The 10 ps difference corresponds to the relative $m/q$ difference of about $3\times 10^{-5}$ for two ion species. Ambiguities may appear in cases when the relative $m/q$ difference is less than $3\times 10^{-5}$. But this uncertainty does not affect the PID for other ions in the same injection. This is important and will be considered in the procedure of $\Delta T$ correction.

\section{$\Delta T$ correction}
As mentioned in the previous sections, the magnetic field instabilities prevent us from determining precise $T$ directly from the acquired raw data, which is a key goal of the experiment. Great efforts have been devoted to remove the $T$ deteriorations caused by magnetic field instabilities. In the following, some techniques previously used in the $\Delta T$ correction are briefly reviewed. Then we present an alternative approach for $\Delta T$ correction based on the successful PID described in subsection~\ref{PID2}.

\subsection{Previous $\Delta T$ correction methods}\label{PMFC}
In early IMS experiments, $T$ from several tens to hundred injections were combined to form a sub-spectrum. Then the correlation-matrix approach~\cite{Audi1986,Radon2000,Yuri2005} was used to obtain the mass values~\cite{Sun2008,Knobel2016A}. One can also combine the sub-spectra into an integrated one via a scaling technique~\cite{ZhangPRL2012}. However, the deterioration of resolution in the sub-spectra due to magnetic field instabilities can not be corrected in these approaches. To avoid this disadvantage, the $T$ spectra relative to several reference nuclides were constructed and used for mass determinations~\cite{TuPL2011,Tu2011NIMA}. Although improvements could be achieved, it is not suitable for most neutron-rich or proton-rich nuclides with low production rates.

Recently, a new method~\cite{shuai-MFC,Zhang2018} has been developed and widely used in the data analysis~\cite{Fu2018,Zhang2018,Xing2018,ZhangPLB2017}. This method assumes that the $T$ \red{peak} for each ion species has a normal distribution and $\Delta T$ due to the variation of magnetic fields is identical for all ions in the same injection. However, the normal distributions of the $T$ peaks can not always be satisfied especially for the ones lying outside the isochronous region. Moreover, $\Delta T$ is not strictly constant since $\Delta T$ depends on ion velocities as indicated in Eq.~(\ref{DeltaT_Xi}).

In the following, we present an alternative approach to achieve a $\Delta T$ correction on an injection-by-injection basis.

\subsection{Definitions of reference frame and common frame}\label{drs}
As discussed in section~\ref{PID2}, PID has been realized for the ions in a single injection. This means that $m/q$ values have been assigned to experimental $T_{exp}$.
We can now plot $T_{exp}$ as a function of $m/q$ (the number of identified particles with well-known masses should be more than two). By taking $B\rho$ and $C$ as free parameters, the scattered data are fitted using Eq.~(\ref{Tsim}), and the set of parameters ($B\rho_{fit},C_{fit}$) are obtained for this injection. Obviously, ($B\rho_{fit},C_{fit}$) is an estimation of the mean orbital length and magnetic rigidity for the ions in this injection. Their spread is mainly caused by the ring's non-zero momentum acceptance. We define $\left\lbrace B\rho_{fit},C_{fit} \right\rbrace$ as the {\it reference frame} for this injection.
\begin{figure} [htb]
	\includegraphics[angle=0,width=8.5 cm]{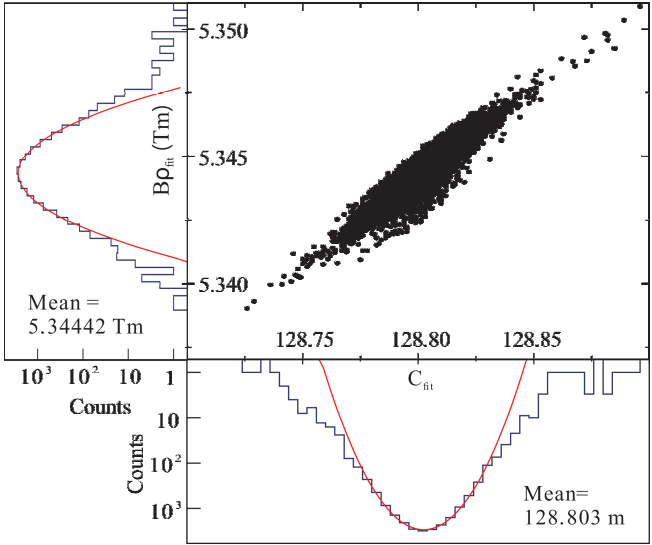}
	\caption{(Colour online) Scatter plot $B\rho_{fit}$ versus $C_{fit}$. The mean values $B\rho=5.34437$ Tm, $C=128.803$ m are obtained. \blue{Adapted from \cite{JLiu}.}
	\label{Brho-C-log2}}
\end{figure}

Fig.~\ref{Brho-C-log2} shows the scatter plot of $B\rho_{fit}$ versus $C_{fit}$ for all injections. The mean values are obtained to be $\overline{B\rho}_{fit}=5.34442$ Tm and $\overline{C}_{fit}=128.803$ m. We note that $\overline{C}_{fit}=128.803$ m agrees well with the orbital length used in the peak $matching $ procedure.
Therefore we define the {\it common frame} as $\left\lbrace B\rho_{cf},C_{cf} \right\rbrace= \left\lbrace 5.34442,128.803 \right\rbrace $. The revolution time correction procedure is in fact the transformation of the experimental $T$ values from their respective {\it reference frames} into the {\it common frame}.

\subsection{$\Delta T$ correction on the injection-by-injection basis}\label{MFCSI}
\begin{figure}[htb]
  	\includegraphics[angle=0,width=8.5 cm]{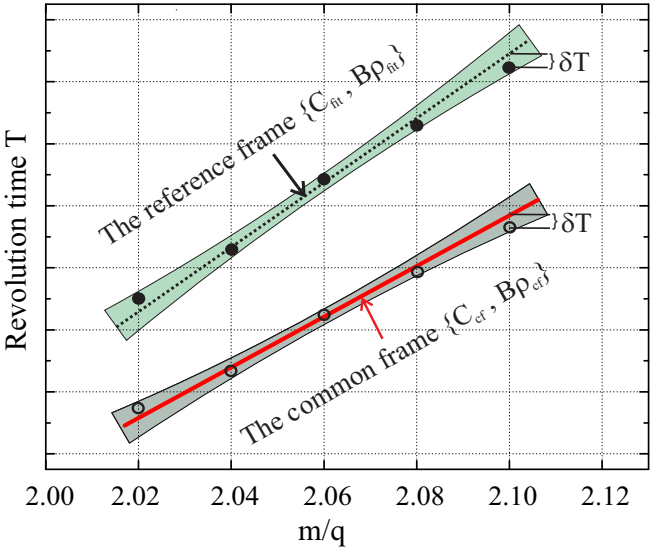}
  	\caption{(Colour online) Schematic illustration for the $\Delta T$ correction. See section~\ref{MFCSI} for details.}
  	\label{Schematic}
  \end{figure}

Fig.~\ref{Schematic} illustrates schematically the principle of our $\Delta T$ correction method. The red solid line in Fig.~\ref{Schematic} is calculated according to Eq.~(\ref{Tsim}) with ($B\rho$,$C$)=($B\rho_{cf}$,$C_{cf}$), describing how $T$ varies with $m/q$ in the {\it common frame} $\left\lbrace B\rho_{cf},C_{cf} \right\rbrace$.

Now suppose a few ions are stored in one injection, and their $T$ are measured. Since the PID has been realized in section~\ref{PID2}, we schematically show the plot of $T_{exp}$ versus $m/q$ in Fig.~\ref{Schematic} as filled circles. By fitting $T_{exp}$ versus $m/q$ using Eq.~(\ref{Tsim}), the parameters ($B\rho_{fit},C_{fit})$ are obtained. The fitted result, $T_{fit}$ versus $m/q$, is shown as the dotted line in Fig.~\ref{Schematic}. Generally, the dotted line is not parallel to the red solid line, and $\delta T=T_{exp}-T_{fit}$ represents the spread of $T_{exp}$ mainly due to the storage ring momentum acceptance.

The ions with uncertain {\it m/q} assignment are not used in the fitting process. These are nuclides with unknown or poorly-known mass, nuclides with possible low-lying unresolvable isomers, or nuclides with $m/q$ values similar to each other.

For the ion with a fixed $m/q$, the $T$ value on the solid line, $T_{cf}$, is in fact the $T$ value when the $reference frame$ translates from $\left\lbrace B\rho_{fit},C_{fit} \right\rbrace $ to $\left\lbrace B\rho_{cf},C_{cf} \right\rbrace $. This can be mathematically described in a first order approximation by the differential formula
   \begin{equation}\label{ADT}
T_{cf}=T_{fit}+\bigg[\frac{C_{cf}-C_{fit}}{C_{fit}}-\frac{1}{\gamma^2}\frac{B\rho_{cf}-B\rho_{fit}}{B\rho_{fit}}\bigg]T_{fit}.
  \end{equation}
Here $T_{fit}$ is determined by $C_{fit}$ and $B\rho_{fit}$ through Eq.~\ref{Tsim}.

Note, that this equation holds only for $T$ values lying on both the dotted and the solid lines in Fig.~\ref{Schematic}.

As $\delta T=T_{exp}-T_{fit}$, we have
\begin{equation}\label{ADT-1}
T_{cf}=T_{exp}-\delta T+\bigg[\frac{C_{cf}-C_{fit}}{C_{fit}}-\frac{1}{\gamma^2}\frac{B\rho_{cf}-B\rho_{fit}}{B\rho_{fit}}\bigg](T_{exp}-\delta T).
 \end{equation}

We define the corrected $T$ as $T_{cor}=T_{cf}+\delta T$, and ignore the influence of $\delta T$ on the second term of Eq.~(\ref{ADT-1}), then $T_{cor}$ is obtained by
\begin{equation}\label{ADT-3}
T_{cor}=T_{exp}+\bigg[\frac{C_{cf}-C_{fit}}{C_{fit}}-\frac{1}{\gamma^2}\frac{B\rho_{cf}-B\rho_{fit}}{B\rho_{fit}}\bigg]T_{exp},
\end{equation}
where $\gamma\simeq1/\sqrt{1-(\frac{1}{v_c}\frac{C_{fit}}{T_{exp}})^2}$. According to Eq.~(\ref{ADT-3}), the $\Delta T$ corrections are made for all ions on the injection-by-injection basis. Finally the corrected $T$ spectrum is obtained.
\begin{figure}[h]
	\includegraphics[angle=0,width=8.5 cm]{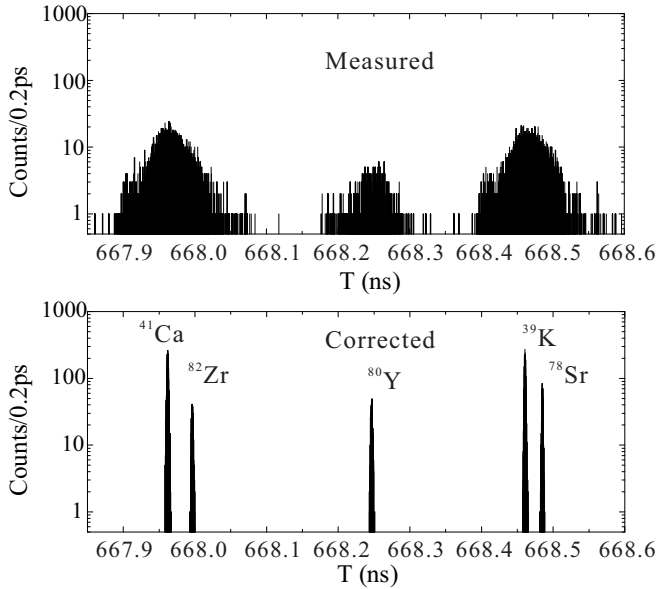}
	\caption{Zoomed $T$ spectrum before and after correction. $^{82}$Zr, $^{80}$Y are not used in the fitting process, but their original $T$ are corrected in the same way as the others \red{by} using Eq.~(\ref{ADT-3}).
		\label{Spectrum_Ori_Cor}}
\end{figure}

The zoomed $T$ spectrum depicting $^{41}$Ca, $^{39}$K, $^{82}$Zr, $^{80}$Y, and $^{78}$Sr is shown in Fig.~\ref{Spectrum_Ori_Cor} where the unresolved peaks in the original $T$ spectrum are clearly separated after the correction procedure.
We note that the $m/q$ ratios of $^{82}$Zr, $^{80}$Y are not used in the fitting process because of the large mass uncertainty of $^{82}$Zr and the known low-lying isomer in $^{80}$Y~\cite{Novikov2001}, but the $\Delta T$ correction for these two nuclides are \red{similarly} made using Eq.~(\ref{ADT-3}).
Taking $^{80}$Y  as an example, the peak width is improved from 43 ps to 2.7 ps after the correction.

\begin{figure}[htb]
	\includegraphics[angle=0,width=8.5 cm]{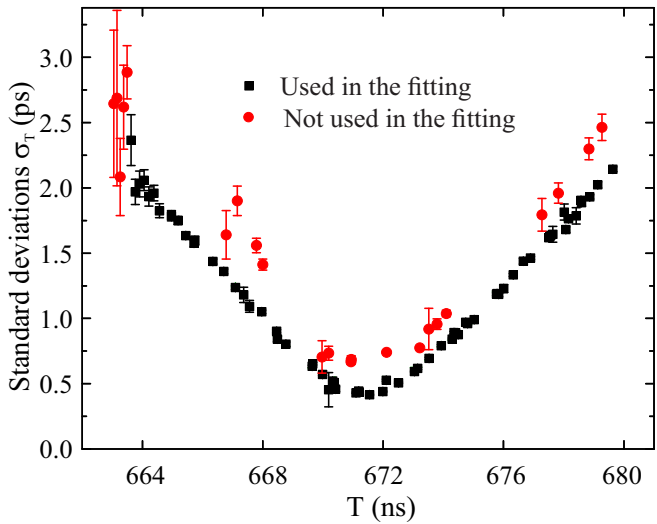}
	\caption{(Colour online) Standard deviations, ($\sigma_T$), deduced from the corrected $T$ spectrum. Ions used in the fitting process are indicated with filled squares, the ions not used in the fitting process are indicated with filled red circles. The nuclides with unresolved isomers are not shown. See text for details.
		\label{result}}
\end{figure}

Fig.~\ref{result} shows the standard deviations ($\sigma_T$) for each $T$ peak extracted from the corrected $T$ spectrum. Several points are worth noting:

(1) The second term of Eq.~(\ref{ADT-3}) reflects the $\Delta T$ shift caused by the variation of magnetic fields. This shift depends on the velocities of stored ions. In our approach, the velocity dependence has been considered.

(2) For the ions not used in the fitting process, for example the unknown-mass nuclides, their $T_{exp}$ have also been corrected in the same procedure. This is because the parameters $B\rho_{fit}$ and $C_{fit}$ are considered to be the approximate $B\rho$ and $C$ values for all ions in the same injection.

(3) The spreads of $T_{exp}$ due to the momentum acceptance of the ring are transferred to $T_{cor}$ after the correction. 
This is schematically indicated by open circles and the gray shadow in Fig.~\ref{Schematic}.
However, if only a very few ions are used in the fitting process, their corrected $T$ spreads will be reduced with respect to the intrinsic ones induced by the finite ring acceptance, while for other ions, not used in the fitting process, the spreads are enlarged. This is demonstrated, respectively, by the black squares and red circles in Fig.~\ref{result}.
It is not consistent with the real situation that $\sigma_T$ should vary coherently along a parabolic-like curve~\cite{Zhang2018,Chen2015}.

In order to avoid this unphysical aberrance in the systematic trends of standard deviations, we used a fixed orbital length in the fitting process, which will be presented and discussed in the next subsection.


\subsection{$\Delta T$ correction with a fixed C}\label{MFCSI-1}

There are two experimental justifications: (1) the momentum distributions of fragments \red{before their injection into the ring} are much larger than the acceptance of the ring; (2) the TOF detector in the ring equipped with a fixed size carbon foil remains stationary during the whole experiment. It is therefore a reasonable assumption that all ions revolve in the ring \red{with the same mean orbit length}.
\begin{figure}[h]
	\includegraphics[angle=0,width=8.5 cm]{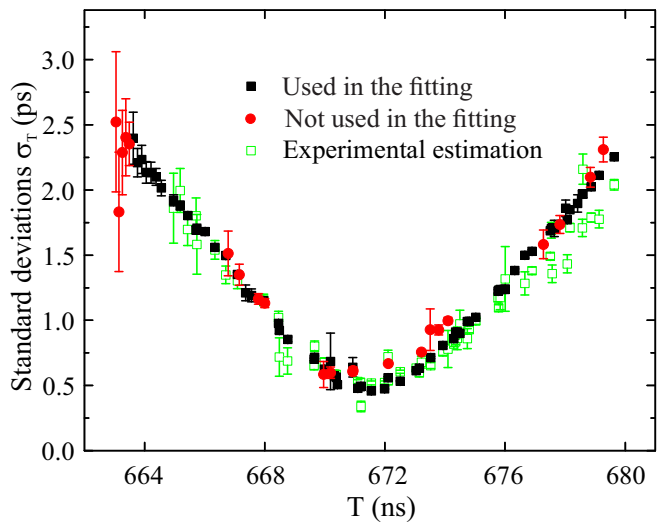}
	\caption{(Colour online) The same as Fig.~\ref{result}, but $C$ is fixed as $C_{cf}$ in the $T$ fitting and correction. Green open squares are intrinsic deviations extracted from injections in which more than two ions of the same species are present. See text for details.
		\label{Cfix}}
\end{figure}

In the following fitting process, we employ a fixed orbital length $C=C_{cf}$. So $C_{fit}=C_{cf}$ and then the $\Delta T$ correction is made according to Eq.~\ref{ADT-3} by
\begin{equation}\label{ADT-4}
T_{cor}=T_{exp}-\frac{1}{\gamma^2}\frac{B\rho_{cf}-B\rho_{fit}}{B\rho_{fit}}T_{exp}.
\end{equation}
As seen in Fig.~\ref{Schematic} and from the derivation of Eq.~(\ref{ADT-3}), the $\Delta T$ correction is a mathematical operation to make the dotted line to match the red solid line. The newly fitted result by using $B\rho=B\rho_{fit}$ and $C=C_{cf}$, i.e., $T_{fit}$ as a function of $m/q$, has still the shape determined by Eq.~(\ref{Tsim}), leading to the mathematical $\Delta T$ correction procedure as formulated by Eq.~(\ref{ADT-4}).

Based on $T_{cor}$ from Eq.~(\ref{ADT-4}), the standard deviation $\sigma_T$ for each ion species is extracted and presented in Fig.~\ref{Cfix}.
We find that the black and red symbols, which represent $\sigma_T$ for ions used and not used in the fitting process, respectively, do nearly overlap.

Independently, one can also estimate the intrinsic $T$ spreads that are not influenced by the field instabilities and the $\Delta T$ correction method. We select injections in which two ions of the same species are present with $T_1$ and $T_2$\red{,} respectively. 
The $T$ difference, $\Delta t=T_1-T_2$, is considered to be only due to the momentum spread of the ions. Furthermore, it directly reflects the momentum acceptance of the storage ring~\cite{Zhang2018}.
Such analyses have been applied to the ion species with high statistics, and their $\Delta t$ spectra have been constructed. The obtained $\sigma(\Delta t)/\sqrt{2}$ values are plotted in Fig.~\ref{Cfix}. One sees that the deduced $\sigma(\Delta t)/\sqrt{2}$ values are in good agreement with $\sigma_T$ obtained from the corrected $T$ spectrum, indicating that the $\Delta T$ shifts due to the magnetic field instabilities have been properly corrected. The mean $T$ and the calculated standard deviations obtained from the correction method are then used for mass determination.
\begin{figure*}\centering
	\includegraphics[angle=0,width=18 cm]{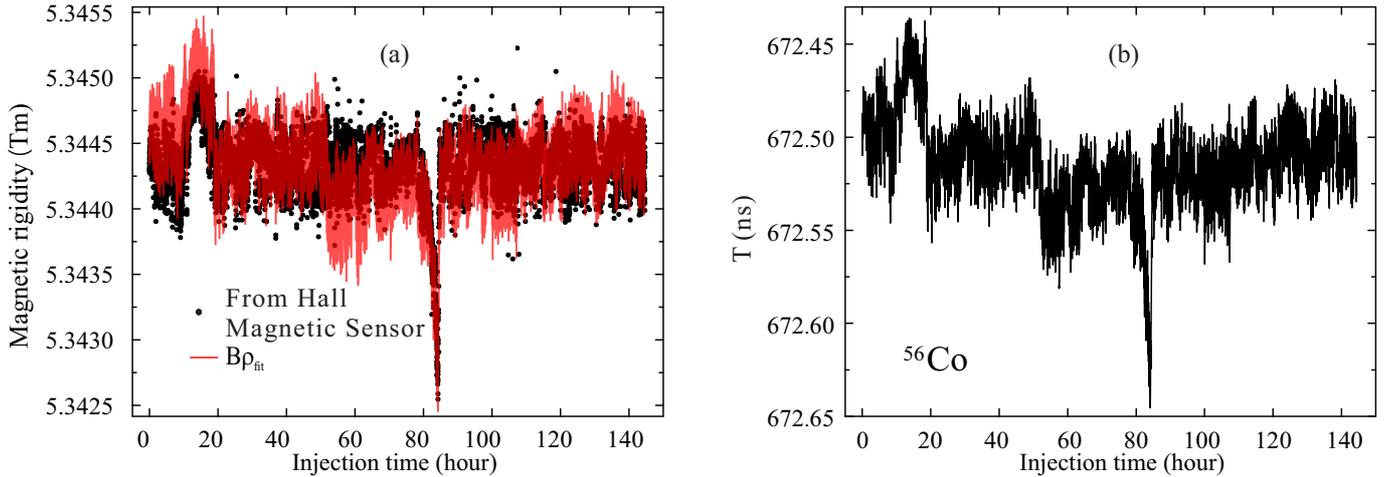}
	\caption{(a): Variation of magnetic rigidity versus injection time. The red line is the fitted $B\rho_{fit}$ with fixed $C$, while black points are deduced from the magnetic field $B$ measured by the Hall probe \red{multiplied by} a constant bending radius \red{of} 6.043 m. (b): The measured $T$ of $^{56}$Co as a function of injection time.
		\label{Brho-Injection-56Co}}
\end{figure*}

Finally we note, that $B\rho_{fit}$ obtained with the fixed $C=C_{cf}$ is a measure of the magnetic rigidity at $C_{cf}$ for each injection. The red line in Fig.~\ref{Brho-Injection-56Co}(a) shows the variation of $B\rho_{fit}$ versus injection time.
Its spread is much smaller than the $B\rho_{fit}$ obtained in section ~\ref{MFCSI} because now the main contribution is due to the instability of magnetic fields ($\thicksim10^{-4}$) and not due to the momentum acceptance($\thicksim10^{-3}$).
The black points are $B\rho$ values deduced from the magnetic field, $B$, measured by a Hall probe and multiplied by a constant $\rho$ value of about 6.043 m (the bending radius is 6.0 m~\cite{XiaNIMA2002}). We find that the deduced magnetic rigidity agrees well with $B\rho_{fit}$. Fig.~\ref{Brho-Injection-56Co}(b) illustrates the measured revolution times, $T_{exp}$, as a function of injection time for $^{56}$Co. Similar variation tendencies are observed for $T_{exp}$, $B\rho_{fit}$, and $B\rho$ deduced from the Hall probe.
This fact indicates that the slow field variation of CSRe can be described approximately by $B\rho_{fit}$ with the fixed $C=C_{cf}$.

\subsection{Mass determination}
After the $\Delta T$ correction, the mass of each nuclide has been re-determined in the same way as described in Ref.~\cite{Zhang2018}. Since some of the mass values~\cite{Xing2018} have already been included into the latest mass table AME$^{\prime}$16~\cite{AME2016}, AME$^{\prime}$12~\cite{AME2012} is used here as the reference basis.
\begin{figure*}[hbt]
	\centering
	\includegraphics[angle=0,width=15.0 cm]{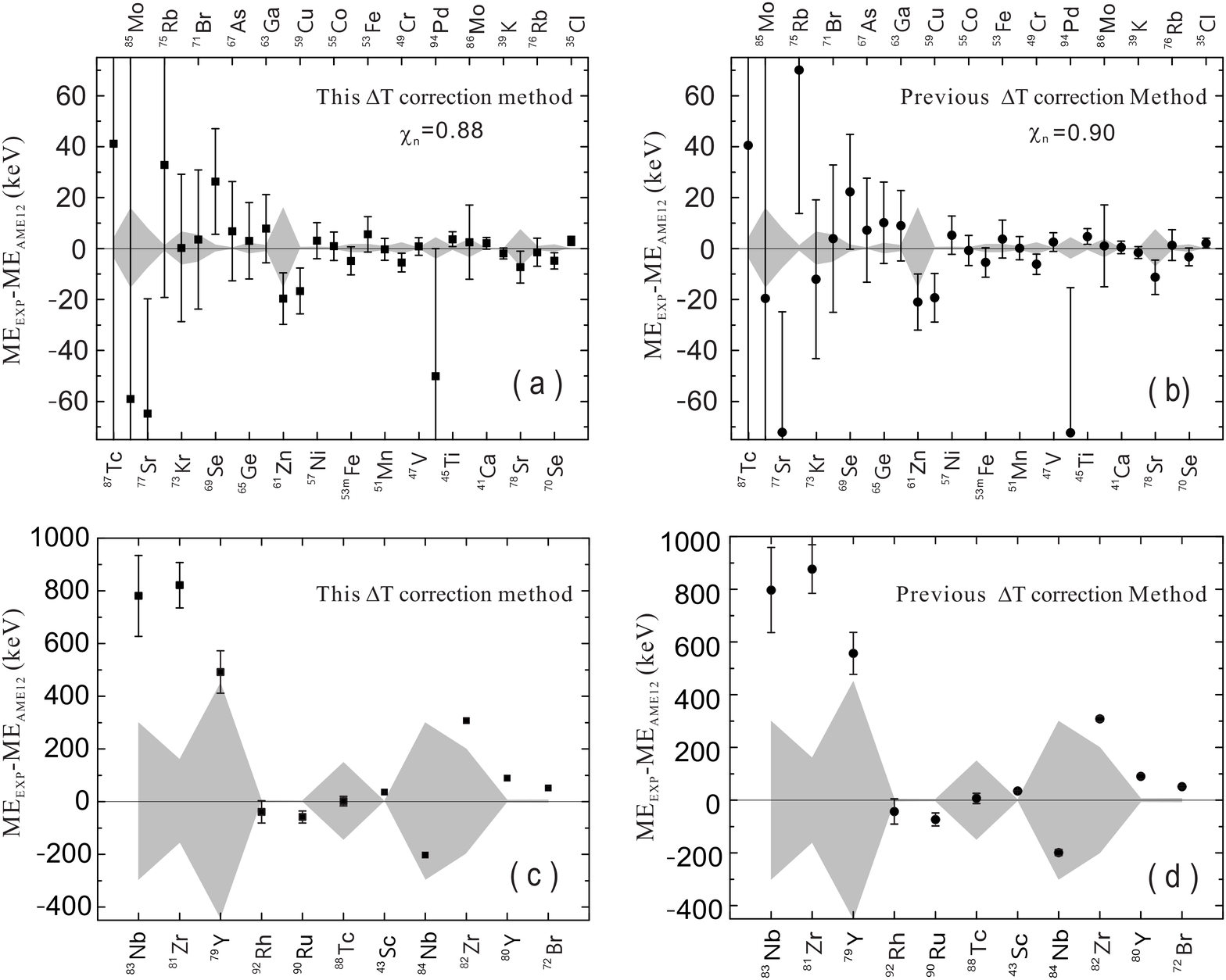}
	\caption{
		(a) and (c): Mass excesses (ME) determined in this work compared with the corresponding literature values. The shadows are the mass uncertainties from AME$^{\prime}$12~\cite{AME2012}. (b) and (d)~\cite{Xing2018}: Same plots as (a) and (c) but based on the previous $\Delta T$ correction method~\cite{shuai-MFC,Zhang2018}. 
		\label{MassComp_Shuai_FixC}}
\end{figure*}

The upper panel of Fig.~\ref{MassComp_Shuai_FixC} shows the mass excess differences between the re-determined masses and those from AME$^{\prime}$12~\cite{AME2012}. Our re-determined masses agree well with the literature ones, confirming the validity of $\Delta T$ correction procedure.
Fig.~\ref{MassComp_Shuai_FixC} also shows the mass determinations by using the previous $\Delta T$ correction method~\cite{shuai-MFC,Zhang2018}. We see that the results from two methods are consistent with nearly the same normalized $\chi_{n}$, which is defined as:
\begin{equation} \label{eq1}
 \chi_{n}=\sqrt{\frac{1}{N_c}\sum_{i=1}^{N_c}\frac{(ME_{{\rm EXP},i}-ME_{{\rm AME12},i})^2}{\sigma_{{\rm EXP},i}^2+\sigma_{{\rm AME12},i}^2}}
 \end{equation}
with $N_c=28$ being the number of nuclides used in the mass re-determination. Both calculated $\chi_{n}=0.88$ and $\chi_{n}=0.90$ are within the expected range of $\chi_{n}=1\pm 1/\sqrt{2N_c}=1\pm 0.13$ at $1\sigma$ confidence level, indicating that no additional systematic errors have to be considered in either case.

In the present experimental data set, we have identified the $T$ peaks corresponding to $^{83,84}$Nb, $^{81,82}$Zr, and $^{79}$Y. Masses of $^{84}$Nb and $^{82}$Zr are unknown, while masses of $^{83}$Nb, $^{81}$Zr, and $^{79}$Y have large uncertainties. Furthermore, we have also observed $^{92}$Rh, $^{88}$Tc, $^{80}$Y, $^{72}$Br, and $^{43}$Sc. Although their ground-state masses are well known, there are low-lying long-lived isomers in these nuclides~\cite{AME2012}. These isomers can be produced in the fragmentation of $^{112}$Sn, but could not be resolved in this experiment. All these nuclides were not used in the fitting process. The masses of these nuclides have been determined according to their corrected $T$ and the $m/q$ versus $T$ calibration using the nuclides with well-known masses. The mass differences with respect to the values in AME$^{\prime }$12~\cite{AME2012} are presented in the lower panel of Fig.\ref{MassComp_Shuai_FixC}. The newly determined masses also agree well with the results in Ref.~\cite{Xing2018}.

There are low-lying long-lived isomers in $^{43}$Sc, $^{80}$Y, and $^{72}$Br at 151-keV, 228-keV, and 100-keV excitation energies~\cite{AME2012}, respectively. The re-determined masses for these three nuclides are slightly larger than the literature ones (see the lower panel of Fig.~\ref{MassComp_Shuai_FixC}). This fact demonstrates the consistency of the present measurements. In addition, there is an unexplained 3$\sigma$ deviation of $^{90}$Ru for the previous $\Delta T$ correction method~\cite{Xing2018}. Based on our new correction method, this discrepancy has been reduced, though is still significant. For more details about the results from the previous $\Delta T$ correction method, the reader is referred to Ref.~\cite{Xing2018}.

We note, that ion velocities have been considered in the present $\Delta T$ correction method. In principle, the $T$ shifts due to the variations of magnetic fields depend on ion velocities (see Eq.~(\ref{DeltaT_Xi})). In the present experiment, $\Delta B/B\sim 10^{-4}$ can contribute 2 ps to the $T$ shifts for the ions at lower and upper limits of the $665-680$ ns $T$ range. This can have a large effect, especially for nuclides with low statistics when a constant $T$ shift for all ions is assumed \red{as} in the previous method~\cite{shuai-MFC,Zhang2018}. Nevertheless, a general agreement of the two methods indicates the validity of the assumption in Refs.~\cite{shuai-MFC,Zhang2018}. A close inspection of Figs.~\ref{MassComp_Shuai_FixC}(a) and \ref{MassComp_Shuai_FixC}(b) reveals that there is still a minor difference between the two approaches: our correction method yields a better agreement with known precise masses than the previous method~\cite{shuai-MFC,Zhang2018}. Most probably this is due to the fact that the velocity dependent effects in the $\Delta T$ correction have been included in the present method.

\section{Summary and conclusions}\label{Summary}
In IMS experiments, the masses of short-lived nuclides are determined on the basis of precise measurements of their revolution times, $T$.
However, the $T$ spectrum is seriously deteriorated by the instabilities of magnetic fields, which make particle identification (PID) difficult and largely reduce mass resolution.
To eliminate this negative effect, we present a $\Delta T$ correction method on injection-by-injection basis.
We firstly achieve a reliable PID in each individual injection, which allows us to use well-known masses for the follow-up $\Delta T$ correction procedure.
An explicit velocity dependence of the correction procedure is derived for the first time in this work.

The PID technique was applied to the data analysis where the $T$ peaks from different ion species overlap due to field instabilities. The technique is also applicable to experiments when hydrogen-like ions are stored in the measurements of neutron-rich fragmentation products, thus making the $T$ peaks overlap.

By defining a {\it common frame} and {\it reference frame} for each individual injection, $\Delta T$ correction has been performed, transforming $T$ values from the individual {\it reference frames} into the {\it common} one. The transformation itself is described by an analytic expression. The velocity dependence of the $T$ shifts due to the field instabilities is automatically considered in the formulations.

We have employed the method to analyze the data from a CSRe experiment based on the fragmentation of $^{112}$Sn beam. The re-determined masses are in good agreement with the literature values, showing that our method can be applied to data even when the magnetic fields of the ring vary at a level of $\Delta B/B\sim 10^{-4}$. Furthermore, our analysis provides mass values consistent with the previous method~\cite{shuai-MFC,Zhang2018} in which a constant $T$ shift for all stored ions in one injection was assumed. While the constant $T$ shift is a good approximation in a limited $T$ range, our new approach can be applied to the entire measured $T$ spectrum.
Last but not least, the method can be conveniently used in on-line data analysis, for example to estimate the production yields of the ions of interest or to monitor the magnetic field and the $\gamma_t$ setting of the ring~\cite{Chen2018}.

The method presented in this work relies on the fact that the momentum distribution of the particles
at the injection into the storage ring is much larger than the injection acceptance itself.
This enables us to assume that the injected ions are distributed stochastically over all available orbits in the ring.
Thus, the mean orbit length of all ions is the same, which is defined by the exact positioning of the TOF detector inside the ring aperture.
This situation is valid for the present operating storage rings ESR at GSI~\cite{ESR}, CSRe at IMP~\cite{CSRe} as well as R3 at RIKEN~\cite{R3}.
The clear advantage of the proposed method is that the injections with very few ions can be used in the analysis.
This is in particular important for measurements of very exotic nuclei produced with very small rates.
In the limiting cases of just 1 or 2 stored ions, it might indeed be impossible to achieve an unambiguous PID.
Such cases need to be inspected separately and may require to combine (several) subsequent injections together.

In future, the IMS experiments are planned in dedicated storage rings, the Collector Ring (CR)~\cite{ILIMA} and the Spectrometer Ring (SRing)~\cite{SRing} being constucted at next-generation radioactive ion-beam facilities FAIR (Germany) and HIAF (China), respectively.
The momentum acceptance of these rings will be dramatically increased~\cite{CR}.
Furthermore, an additional information on the velocity of each stored particle will become available~\cite{2TOF1,2TOF2,2TOF3}.
Therefore, the method proposed in this work will need to be further improved.

\section*{Acknowledgments}
We thank the staffs in the accelerator division of IMP for providing stable beam. This work is supported in part by the National Key R\&D Program of China (Grant No.2018YFA0404401 and No.2016YFA0400504), the Key Research Program of Frontier Sciences of CAS (Grant No. QYZDJ-SSW-SLH005), the Key Research Program of the Chinese Academy of Sciences (Grant No. XDPB09), the NSFC (Grants No. 11605248, No. 11605249, No. 11605252, No. 11605267), the European Research Council (ERC-CG) under the EU Horizon 2020 research and innovation programme (ERC-CG 682841 "ASTRUm"). Y.M.X acknowledges the support from CAS "Light of West China" Program. Y.H.Z. acknowledges support by the ExtreMe Matter Institute EMMI at the GSI Helmholtzzentrum f{\"u}r Schwerionenforschung, Darmstadt, Germany. Y.A.L. is supported by the CAS visiting professorship for senior international scientists (Grant No. 2009J2-23), the CAS External Cooperation Program (Grant No. GJHZ1305), and the HGF-CAS joint research group (Grant No. HCJRG-108).

\end{document}